\documentclass[a4paper,fleqn]{cas-dc}

\usepackage[switch]{lineno}

\usepackage[authoryear]{natbib}
\newcommand{\araa}{ARA\&A}
\newcommand{\aj}{AJ}
\newcommand{\apj}{ApJ}
\newcommand{\apjl}{ApJ}

\newcommand{\mnras}{MNRAS}
\newcommand{\aap}{A\&A}

\newcommand{\icarus}{Icarus}

\newcommand{\nat}{Nature}

\def\tsc#1{\csdef{#1}{\textsc{\lowercase{#1}}\xspace}}
\tsc{WGM}
\tsc{QE}

\begin{document}
\let\WriteBookmarks\relax
\def\floatpagepagefraction{1}
\def\textpagefraction{.001}
\shorttitle{Planet Nine's influence on Distant TNO inclinations}  
\shortauthors{Bansal et al.}
\title [mode = title]{Distant TNO Inclinations as a Constraint on Primordial Cluster Perturbations in the Presence of Planet Nine}

%

\author[1]{Avni Bansal}[orcid=0009-0002-9042-479X]

\ead{abansal3@caltech.edu}

\affiliation[1]{
            addressline={Cahill Center for Astrophysics, California Institute of Technology, 1216 East California Boulevard}, 
            city={Pasadena},
            postcode={91125}, 
            state={CA},
            country={USA}}

\author[2]{Ian Brunton}

\affiliation[2]{
            addressline={Division of Geological and Planetary Sciences, California Institute of Technology}, 
            city={Pasadena},
            postcode={91125}, 
            state={CA},
            country={USA}}

\author[2]{Konstantin Batygin}

\author[3]{Gabriele Pichierri}

\affiliation[3]{
            addressline={Dipartimento di Fisica, Universit\`{a} degli Studi di Milano, Via Celoria 16}, 
            city={20133 Milano},
            country={Italy}}

\author[4]{David Nesvorn\'{y}}

\affiliation[4]{
            addressline={Department of Space Studies, Southwest Research Institute, 1301 Walnut St., Suite 400}, 
            city={Boulder},
            postcode={80302},
            state={CO},
            country={USA}}

\begin{abstract}
The Sun was almost certainly born in a stellar cluster, implying some degree of external forcing from neighboring stars during the Solar System's infancy. Published estimates of the strongest relevant stellar encounter, however, span a broad range, from relatively gentle perturbations to violently disruptive flybys. The modest inclination dispersion of the distant trans-Neptunian population has previously been used to argue that strong primordial encounters were unlikely and that the outer Solar System was not violently stirred at birth. In this work, we independently examine this constraint and ask whether this conclusion holds in the presence of Planet Nine. Because Planet Nine can reshape the distant trans-Neptunian population over gigayear timescales, the extent to which the present-day inclination distribution preserves a direct record of primordial cluster excitation is a priori unclear. To address this question, we compare long-term $N$-body integrations in which distant test particles begin either in strongly cluster-perturbed configurations or in comparatively quiescent ones, and are then evolved under the influence of the known giant planets, the Galactic tide, the effects of passing stars, and Planet Nine. We find that Planet Nine does not transform a highly excited primordial population into one resembling the observed low-inclination distant sample. Instead, comparisons between the simulated and the observed distant TNO populations indicate that the observed distant TNOs are most consistent with a relatively mildly perturbed birth environment, even in the presence of Planet Nine. If this comparatively low inclination dispersion is confirmed to be an intrinsic feature of the distant TNO population, rather than an artifact of observational bias, it constitutes a robust constraint on the severity of primordial cluster perturbations.
\end{abstract}

\begin{keywords}
Planet Nine \sep Cluster Perturbations \sep Distant Trans-Neptunian Objects
\end{keywords}

\maketitle

\section{Introduction}\label{}
Over the past three decades, the discovery and detailed characterization of the Kuiper belt has transformed the distant Solar System from a speculative population into a quantitative record of post–nebular evolution \citep{morby_kbreview_2008, nesvorny2018dynamical}. Observations of the Kuiper belt have motivated a coherent dynamical narrative linking an early epoch of planet–disk interaction and subsequent reshaping of the outer Solar System to the present day outer Solar System \citep[e.g.,][]{Tsiganis_2005, Levinson_2008_GPM, Morb_2009, Nesvorny_2015}. Accordingly, today the orbital architecture of trans-Neptunian small bodies provides some of the most stringent constraints on the timing and character of giant-planet migration and instability.

The diagnostic power of the Kuiper belt will sharpen substantially in the coming years, as wide-field surveys like the newly completed Vera C. Rubin Observatory continue to map the distant reaches of the outer solar system \citep{Ivezic_verarubin}. With better characterized selection-functions and increased sensitivity, this next generation of observations will be able to probe the weakly bound populations of objects in the outer scattered disk and the inner Oort cloud. These populations would have been the most sensitive to external perturbations operating during the cluster phase of Solar System formation, and so are expected to act as tracers of the Solar System's early environment \citep{ADAMS_2010_review}.

Some level of primordial external forcing on the distant population is unavoidable. Surveys of nearby star-forming regions indicate that Sun-like stars predominantly form in clustered environments, so close stellar passages and the time-varying tidal field of the natal cluster likely acted on the young Solar System \citep{Lada_2003, Porras_2003}. Meteoritic records of short-lived radioisotopes suggest enrichment from nearby massive stars, again pointing to a birth environment in which external interactions could occur \citep{Cameron_1977}. Yet while the existence of perturbations is well motivated, the frequency of close encounters, the duration of cluster membership, and the strength of the cluster tide are all debated. The exact degree to which the present-day outer Solar System should bear the imprint of its birth environment is not known.

A stringent empirical constraint on any external excitation is provided by the cold classical Kuiper belt. Because this population is believed to have formed largely in situ \citep{batygin2011retention, nesvorny2019trans}, its narrow inclination distribution (inclination dispersion $1.7^\circ$) is naturally interpreted as a fragile relic sensitive to perturbations. Sufficiently close stellar perturbations can heat this component of the Kuiper Belt and raise its inclination dispersion, so the observed coldness of the classical belt has long been viewed as an avenue towards placing upper bounds on the intensity of any cluster-driven excitation. \citet{Batygin_2020} quantified this property, placing an upper bound on the sun's birth environment of the form $\chi\lesssim 2-3 \times 10^4 \, \text{Myr/pc}^3$, where $\chi = \int n(t)dt$ and $n(t)$ is the instantaneous
number density of the immediate solar system environment. \citet{Siraj_2025} corroborated this bound with numerical simulations, and we further confirm the bound with an expanded suite of simulations in the Appendix.

More recently, it has been argued that even tighter bounds on $\chi$ may be obtained by considering the orbital inclination dispersion of the most distant trans-Neptunian objects. Although much broader than the cold belt, this population is also observed to have a relatively narrow inclination dispersion. If this is truly the case, then strong cluster perturbations would be disfavored, potentially by up to an order of magnitude relative to limits inferred from the cold classical belt \citep{hu2025earlystellarflybysunlikely}. This line of reasoning, however, rests on two assumptions: that survey biases against high-inclination distant objects are not severe, and that the contemporary distant population is a faithful tracer of the primordial excitation from the cluster phase.

The latter assumption is especially nontrivial in light of the hypothesized Planet Nine (see \citealt{batygin2019planet} for a review). The existence of Planet Nine has been invoked to account for several apparent anomalies in the outer Solar System, including clustering in longitudes of perihelion and orbital-plane orientations among extreme TNOs, the delivery of low-inclination Neptune-crossers \citep{Batygin_2024}, and the production of high-inclination and retrograde Centaur-like objects at large semimajor axes (as well as the production of their lower--semimajor-axis counterparts; \citealt{BatyginandBrown2016a, BatyginandBrown2016b}). Irrespective of the ultimate interpretation of these observational signatures, a key dynamical point is clear: if Planet Nine exists, its secular perturbations must unavoidably modulate the eccentricity and inclination distribution of distant small bodies over gigayear timescales. Indeed, some degree of inclination pumping is not merely incidental but essential to the Planet Nine mechanism, since the production of the most highly inclined and eccentric objects relies on sustained eccentricity–inclination coupling driven by the planet’s gravitational influence \citep{batygin2017dynamical, bkr_18}. This interaction acts to lift the perihelia of momentarily-inclined TNOs, potentially removing them from the observable domain. 

This raises a basic question -- could a Solar System that emerged from a strongly perturbative birth environment subsequently evolve, under the action of Planet Nine, into a state whose distant population appears to exhibit only modest inclination dispersion today (after accounting for observational bias)? Or does the existence of a low-inclination distant population require that the outer Solar System began in a comparatively cold configuration even before Planet Nine’s long-term sculpting? In this paper, we address this question with a suite of numerical experiments that incorporate Planet Nine and follow the long-term evolution of distant trans-Neptunian test populations with initial conditions either heavily perturbed by cluster-excitation, or relatively unperturbed by the solar system's birth environment. Our aim is to assess whether Planet Nine, acting in conjunction with observation biases, can dynamically “cool” the observable portion of a highly dispersed primordial population into something resembling the observed low dispersion, or whether the observed inclination structure of the distant Solar System necessarily implies mild primordial perturbations even in the presence of Planet Nine. It is worth noting that the survival of Planet Nine in face of cluster-induced perturbations is not guaranteed. \citet{Li_2016} have carried out a detailed numerical investigation of this and have shown that Planet Nine is likely to be ejected if the Sun resides within its birth cluster longer than $\Delta t \gtrsim100$ Myr. We assume this has not yet happened.

The remainder of the manuscript is organized as follows: Section \ref{sec:sims} describes how our simulations are constructed. Section \ref{sec:results} compares the inclination and perihelion distributions in our simulations with the observed distributions. In Section \ref{sec:Discussion} we discuss the extent to which primordial TNOs were perturbed by the birth environment of the solar system, and in Section \ref{sec:Conclusion} we conclude. Cumulatively, our results indicate that while Planet Nine cannot cool highly dispersed populations to anything consistent with the observed low dispersions, its presence simultaneously does not overexcite the distant population, suggesting that the observed low dispersion is primordial.

\section{Simulation Set-Ups} \label{sec:sims}
Our suite of N-body integrations was designed to quantify how Planet Nine modifies the inclination and eccentricity architecture of the distant trans-Neptunian region, and elucidate the extent to which present-day orbital dispersions can be mapped back to primordial excitation.

\subsection{Cluster-Influenced Simulations}
 Our initial test-particle ensemble, representing the population of trans-Neptunian minor bodies, was drawn from the synthetic distant population produced in the numerical experiments of \citet{Nesvorny2023}. \citet{Nesvorny2023} carried out $N$-body simulations of the giant planets and a large population of disk planetesimals, tracking Neptune's outward migration and instability while incorporating the gravitational potential of the Sun's birth cluster, encounters with passing cluster stars, the Galactic tide, and field-star encounters. In practice, we drew test particles from the $\tau = 300 \, \mathrm{Myr}$ timestamp ephemeris of the \citet{Nesvorny2023} \texttt{cluster\_2} simulation. $\tau = 300 \, \mathrm{Myr}$ is late enough that it is after the dispersal of the natal cluster and the completion of the main phase of giant-planet migration and instability. We chose the \texttt{cluster\_2} simulation because it includes a strong variant of cluster effects with close stellar encounters at the limit of the cluster perturbation allowed by the cold belt. This choice ensures that the starting distribution is physically motivated and already contains the maximal allowable excitation imparted during the early Solar System’s formative epoch. From this parent distribution, we selected $10000$ objects in the distance regime relevant to Planet Nine dynamics (semi-major axis $\sim100-5000$ AU ), and treated all small bodies as massless test particles. Standard cuts were applied to exclude strongly Neptune-crossing objects at the starting epoch (perihelia $< 30$ AU).

To reduce computational cost while retaining the essential long-term dynamics, only Neptune and Planet Nine were included as active, fully interacting perturbers. The effects of Jupiter, Saturn, and Uranus were captured through an effective quadrupole moment of the Sun. Specifically, we represented the averaged secular precession that the inner giant planets impose on distant orbits by augmenting the central potential with a solar $J_2$-like term calibrated to reproduce the appropriate apsidal and nodal precession rates (see e.g., \citealt{batygin2019planet}). This approximation preserves the key low-frequency structure of the problem that governs secular coupling in the distant Solar System, while avoiding the expense of explicitly integrating all four giant planets over gigayears.

Planet Nine was treated as an additional massive body with fixed mass and initial orbital elements for a given run. Planet Nine parameters were drawn from $m_9 \in \{5, 7.07, 10\} \, M_\oplus$, $i_9 = 20^\circ$ and $e_9 \in \{0.2, 0.35, 0.5\}$. For each $e_9$, the semi-major axis $a_9$ was chosen such that the perihelion distance stayed approximately fixed at $q_9 \sim [250,300]$ AU. This parameter space approximately follows the parameter space outlined in \citet{batygin2019planet}, consistent with what is typically invoked in Planet Nine studies. We note that we only use one inclination value ($i_9 = 20^\circ$) because the observationally allowed range is narrow ($i_9 \sim 15$--$25^\circ$), so varying $i_9$ within this range is unlikely to substantially alter the results.

External forcing from the Solar System’s environment was included via two ingredients. First, we incorporated the Galactic tide using a standard prescription appropriate for the solar neighborhood (see \citealt{Nesvorny_2017} and references therein). This component is chiefly relevant for particles that diffuse to very large semi-major axes ($a \sim 2000-5000$ AU), where tidal torques can compete with planetary perturbations and modify perihelia on long timescales \citep{Saillenfest_2020}. Second, passing stars were modeled using the classical impulse approximation \citep{HeislerTremaine1986}. Each passing star's velocity was set to $\sqrt{2} \times 1 \, \mathrm{km \, s}^{-1}$, corresponding to the RMS velocity expected for isotropic random motion in stellar clusters like the Sun's birth environment \citep{ADAMS_2010_review, Lynne}. This differs from treatments that explicitly integrate the stellar trajectory as an additional gravitating body during the flyby. We note that the impulse approximation starts to break down and does not strictly apply for very low semi-major axes ($a \sim 100$ AU) because the encounter time and orbital period are comparable there. However, the perturbation imparted by a flyby is the smallest in the low semi-major axis regime, so the exact dynamics of perturbations in this regime do not significantly affect the inclination dispersion we are interested in. Additionally, we verified through direct comparisons that the impulse treatment reproduces the same statistical outcomes as explicit modeling for the encounter regions relevant here, to within the level that matters for our conclusions. 

Simulations with initial conditions excited by the cluster, as described here, are henceforth referred to as cluster-influenced simulations.

\subsection{Integration}
All integrations were carried out with the \texttt{MERCURY6} gravitational dynamics software package \citep{Chambers_1999} using a conservative Bulirsch--Stoer algorithm. For each Planet Nine configuration, we evolved the selected test-particle ensemble for $4\;\mathrm{Gyr}$, and treated the particle distribution across the final Gyr of the integrations as a quasi steady-state distribution. Output was recorded at regular intervals of $2.5\;\mathrm{Myr}$ in order to construct the inclination and eccentricity statistics and identify the subset of objects that remain in the observationally relevant region at the end of each run.

\subsection{Cluster-Free Simulations}
For the control case of initial conditions corresponding to no cluster perturbations, we draw on results already published in the literature. In the simulations of \citet{batygin2019planet}, the trans-Neptunian minor body population is represented by $1000$ test particles with inclinations drawn from a half-Gaussian distribution with a dispersion of $\sigma_i = 15^\circ$ and $a \in (100,800)\;AU$. Strongly Neptune-crossing objects are excluded by selecting $q \in (30,100)\;$AU, and integration is carried out in a manner similar to the cluster-influenced case.

We consider eight instances of these simulations, with Planet Nine parameters $5\,M_{\oplus},\, a_9 \in \{400, 500\}\,\text{AU},$ $e_9 \in \{0.25, 0.35, 0.45, 0.55\}$ and $i_9=20^\circ$. Unlike the cluster-influenced runs, the perihelion is not fixed in these simulations. The semi-major axis and eccentricity values sampled here differ slightly from those used for the cluster-influenced runs because the two sets of initial conditions were constructed independently, with the cluster-free runs following the parameter set of \citet{batygin2019planet}. This difference does not affect our conclusions, since the cluster-free runs serve only to demonstrate that Planet Nine does not overexcite an initially cold distant population, for which a representative range of $e_9$ is sufficient.

Eight simulations are sufficient for our purposes here, since we are only interested in showing that Planet Nine does not overexcite distant objects in the absence of strong cluster effects. In fact, even one simulation where Planet Nine does not overexcite the distant objects would be sufficient to show that the Planet Nine scenario is not necessarily inconsistent with an unexcited distant population in the absence of a cluster. A broader set of Planet Nine simulations with initial conditions not affected by cluster perturbations (henceforth referred to as cluster-free simulations) are explored in detail in \citet{brobat2021}.

\section{Results}\label{sec:results}

\begin{table}
\centering
\caption{Best-fit width $w$ of the inclination distribution of the high-$q$ TNOs from Planet Nine simulations with cluster-influenced and cluster-free initial conditions. The width is estimated by maximizing the log-likelihood under $f(i) = \sin i \, \exp(-i^2/2w^2)$. Uncertainties are $\lesssim 0.1^\circ$ due to large sample sizes ($N \sim 10^5$). Also shown is the von Mises concentration parameter $\kappa$ of the longitude of perihelion distribution.}
\label{tab:sim_sigma}
\small
\begin{tabular}{cccccc}
\hline
$m_9$ & $a_9$ & $i_9$ & $e_9$ & $w$ & $\kappa$ \\
($M_\oplus$) & (AU) & ($^\circ$) & & ($^\circ$) & \\
\hline
\multicolumn{6}{c}{\textit{Cluster-influenced}} \\
\hline
5.00 & 367 & 20 & 0.2 & 26.5 & 0.261 \\
5.00 & 420 & 20 & 0.35 & 26.0 & 0.311 \\
5.00 & 480 & 20 & 0.50 & 27.0 & 0.487 \\
7.07 & 356 & 20 & 0.2 & 27.0 & 0.338 \\
7.07 & 433 & 20 & 0.35 & 27.0 & 0.439 \\
7.07 & 497 & 20 & 0.50 & 26.0 & 0.593 \\
10.00 & 356 & 20 & 0.2 & 27.0 & 0.464 \\
10.00 & 433 & 20 & 0.35 & 27.0 & 0.690 \\
10.00 & 540 & 20 & 0.50 & 27.5 & 0.711 \\
\hline
\multicolumn{6}{c}{\textit{Cluster-free}} \\
\hline
5.0 & 400 & 20 & 0.25 & 16.0 & 2.880 \\
5.0 & 400 & 20 & 0.35 & 17.5 & 2.501 \\
5.0 & 400 & 20 & 0.45 & 17.0 & 2.209 \\
5.0 & 400 & 20 & 0.55 & 18.5 & 2.682 \\
5.0 & 500 & 20 & 0.25 & 17.0 & 1.193 \\
5.0 & 500 & 20 & 0.35 & 18.5 & 1.786 \\
5.0 & 500 & 20 & 0.45 & 18.0 & 3.130 \\
5.0 & 500 & 20 & 0.55 & 19.5 & 3.101 \\
\hline
\end{tabular}
\end{table}
\subsection{Comparing simulation and observed inclination distributions}
\label{section}
To assess whether Planet Nine can confine initially excited TNOs to their observed inclination distribution, we need to know what the observed inclination distribution is. The inclinations of the discovered Kuiper belt objects (KBOs) are generally biased towards the ecliptic latitude of observation, so we apply the debiasing procedure developed in \citet{Brown_2001} to infer the intrinsic inclination distribution as follows:

Let $f_e(i)$ be the distribution of inclinations of KBOs near the ecliptic. For a single dynamical population (here the high-$q$ TNOs), a natural functional form for $f_e(i)$ is a half-Gaussian. Using a simple geometric argument, it can be shown that the intrinsic inclination distribution of the full sample, $f_t(i)$, is related to the ecliptic distribution by $f_t(i) = \sin(i) \cdot f_e(i) = \sin(i) \cdot A \, e^{-i^2 / 2w^2}$. Thus, the problem of determining $f_t$ is reduced to the problem of determining $w$ ($A$ is just a normalization constant). We can find $w$ by calculating the probability that an object $j$, discovered at some latitude $\beta_j$, has an inclination less than or equal to its observed inclination $i_j$. This probability is given by:
\begin{align*}
P_j = \int_{\beta_j}^{i_j} \frac{f_t(i')}{(\sin^2 i' - \sin^2 \beta_j)^{1/2}} \, di' \\
\times \left[ \int_{\beta_j}^{\pi/2} \frac{f_t(i')}{(\sin^2 i' - \sin^2 \beta_j)^{1/2}} \, di' \right]^{-1}.
\end{align*}
If $f_t$ represents the correct intrinsic inclination distribution, we expect $P_j$ to be uniform between $0$ and $1$. We can evaluate $P_j$ for all discovered high-$q$ TNOs with different values of $w$. Then, we can check how close $P_j$ is to uniform using the $D\sqrt{N}$ statistic from the Kuiper modification of the KS test, as a function of $w$. The $w$ value that results in $P_j$ being closest to uniform must be the intrinsic $w$.

\begin{figure}
\centering
\includegraphics[width=\columnwidth]{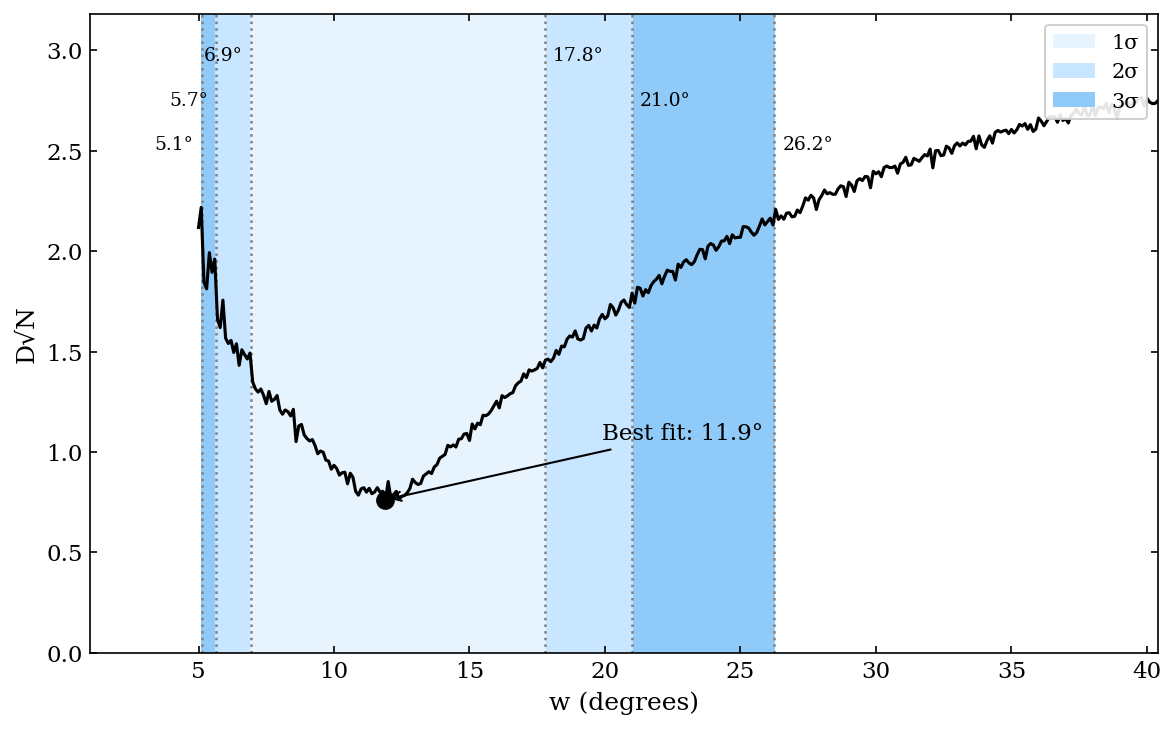}
\caption{$D\sqrt{N}$ statistic as a function of the inclination distribution width $w$. The best-fit value and $1\sigma,2\sigma$ and $3\sigma$ confidence intervals are indicated. The $1\sigma$ confidence interval spans the range of $w$ such that $D\sqrt{N}$ remains below the calibrated Kuiper test threshold at the 84.1\% level, following \citet{Brown_2001}.}
\label{fig:dsqrtn}
\end{figure}

\renewcommand{\thefootnote}{\fnsymbol{footnote}}
For this analysis, we selected all TNOs with $q \in (40, 80) \, \mathrm{AU}$ and $a \in (200, 2000) \, \mathrm{AU}$ from the Minor Planet Center database. After excluding objects with large orbital uncertainties, our sample contained $19$ high-$q$ TNOs\footnote[2]{The selected TNOs are 2000 CR105, Alicanto, 2013 UT15, Leleakuhonua, 2013 RA109, Sedna, 2010 GB174, 2013 FT28, 2013 SY99, 2014 SR349, 2014 WB556, 2015 KG163, 2015 RX245, 2016 SD106, 2017 OF201, 2018 VM35, 2021 RR205, 2023 KQ14, and 2024 FX26.}.
\renewcommand{\thefootnote}{\arabic{footnote}} All selected objects have inclinations less than $40^\circ$. We computed the mean orbital pole of these objects and applied a Rodrigues rotation to transform to a coordinate system where the mean pole defines the $z$-axis, and recomputed inclinations in this rotated frame. Applying the procedure, we find $w_{obs} = 12^\circ{}^{+6}_{-5} $ (see Figure~\ref{fig:dsqrtn}). 

To compare with simulations, we selected simulated particles satisfying the same orbital cuts ($a \in (200, 2000) \, \mathrm{AU}$ and $q \in (40, 80) \, \mathrm{AU}$) and $i < 40^\circ$ at $10 \, \mathrm{Myr}$ intervals over the last Gyr of evolution. We computed the mean pole and applied the same coordinate transformation as for the observed sample, and then determined the maximum likelihood estimate of $w$ by calculating log-likelihoods as a function of $w$ (see Table~\ref{tab:sim_sigma}). 

\begin{figure}
\centering
\includegraphics[width=\columnwidth]{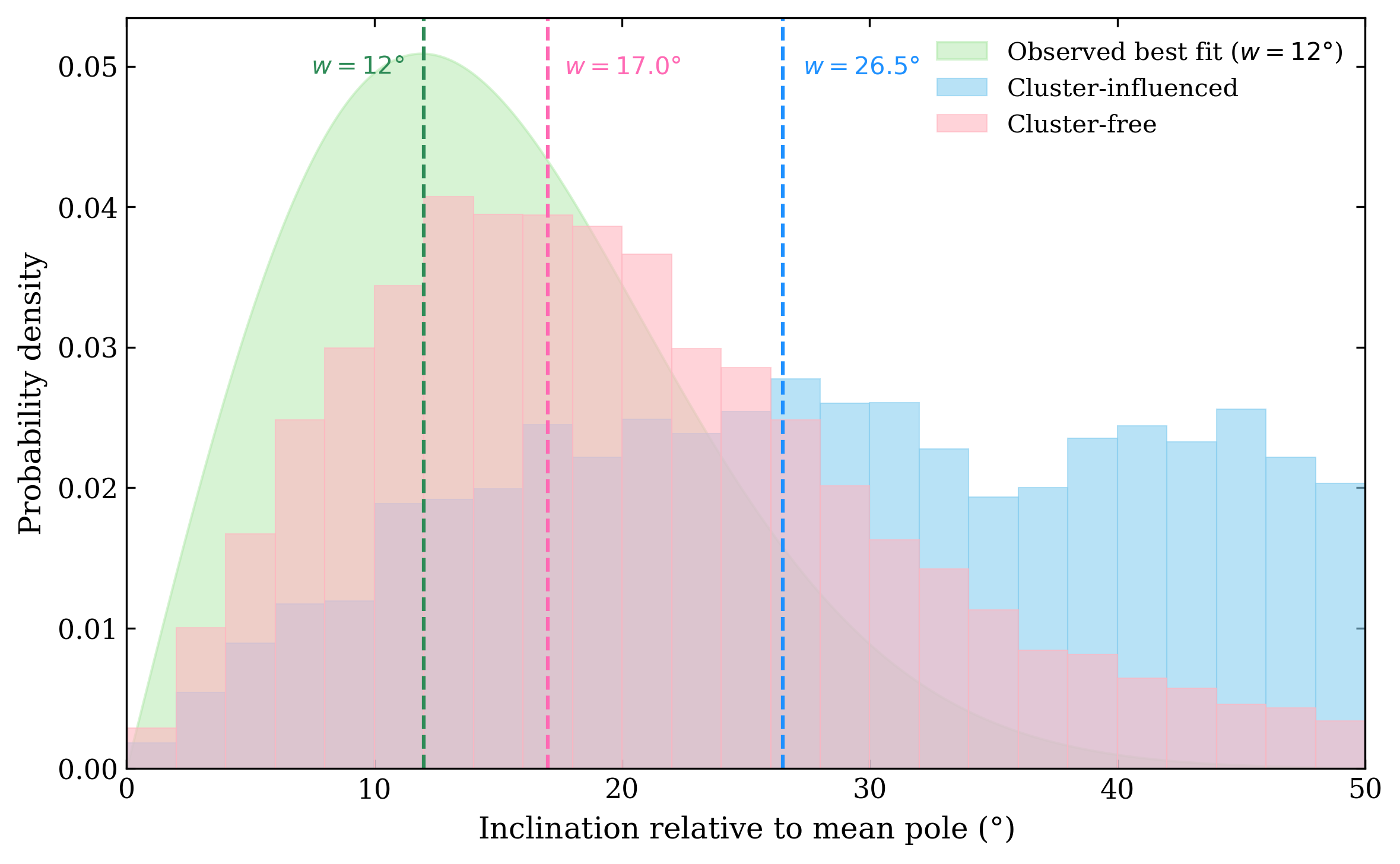}
\caption{Distribution of inclinations relative to the mean orbital pole for high-$q$ TNOs in Planet Nine simulations with cluster-influenced (blue; $m_9 = 5\,M_\oplus$, $e_9 = 0.2$, $i_9 = 20^\circ$) and cluster-free (pink; $m_9 = 5\,M_\oplus$, $e_9 = 0.45$, $i_9 = 20^\circ$) initial conditions, and the best-fit intrinsic inclination distribution ($f_t (i)$) for the 19 high-q observed TNOs. Dashed lines indicate the best-fit width w.}
\label{fig:inc_histogram}
\end{figure}

We find that Planet Nine is unable to confine the distant objects in the cluster-influenced simulations. The inclination dispersion is $\sim 26-27.5^\circ$ across cluster-influenced simulations, with small dependence on Planet Nine mass (see Table \ref{tab:sim_sigma}). The reason becomes apparent if we look at figure \ref{fig:inc_histogram}. The inclination distribution for cluster-influenced simulations is visibly flat for $i \gtrsim 17^\circ$, rather than the sine-half-gaussian form we assumed. The flatness of the distribution reflects a population that was isotropically scattered by strong cluster perturbations and subsequently perturbed but not vertically confined by Planet Nine. Using the same Monte Carlo calibrated Kuiper test from \ref{section}, a width $w = 26^\circ$ is rejected at the $\sim 3\sigma$ level ($p = 0.0014$).

Meanwhile the cluster-free simulations have $w \sim16-19^\circ$. While this is slightly higher than the inferred intrinsic $w_{obs} = 12^\circ$, four out of the eight simulations have dispersions within $1\sigma$ of the observed dispersion, and all are within $1.5\sigma$. The relatively low dispersions tell us that the dispersion of the cluster-free simulations stay very close to the primordial dispersion, which we had set to $15^\circ$. This indicates that Planet Nine does not appreciably heat the distant objects. It is possible that initializing the distant objects closer to $12^\circ$ may result in final dispersions that are also closer to $12^\circ$. However, as we note in the discussion, the small number of known distant TNOs and a number of well-known biases affecting inclinations mean that there is uncertainty in our intrinsic $w_{obs}$ estimate. Given current sample sizes used to infer $w_{obs}$, as well as the presence of biases, we do not regard the mild discrepancy between the cluster-free and observed dispersions as significant.
\subsection{Perihelion clustering}

\begin{figure}
\centering
\includegraphics[width=\columnwidth]{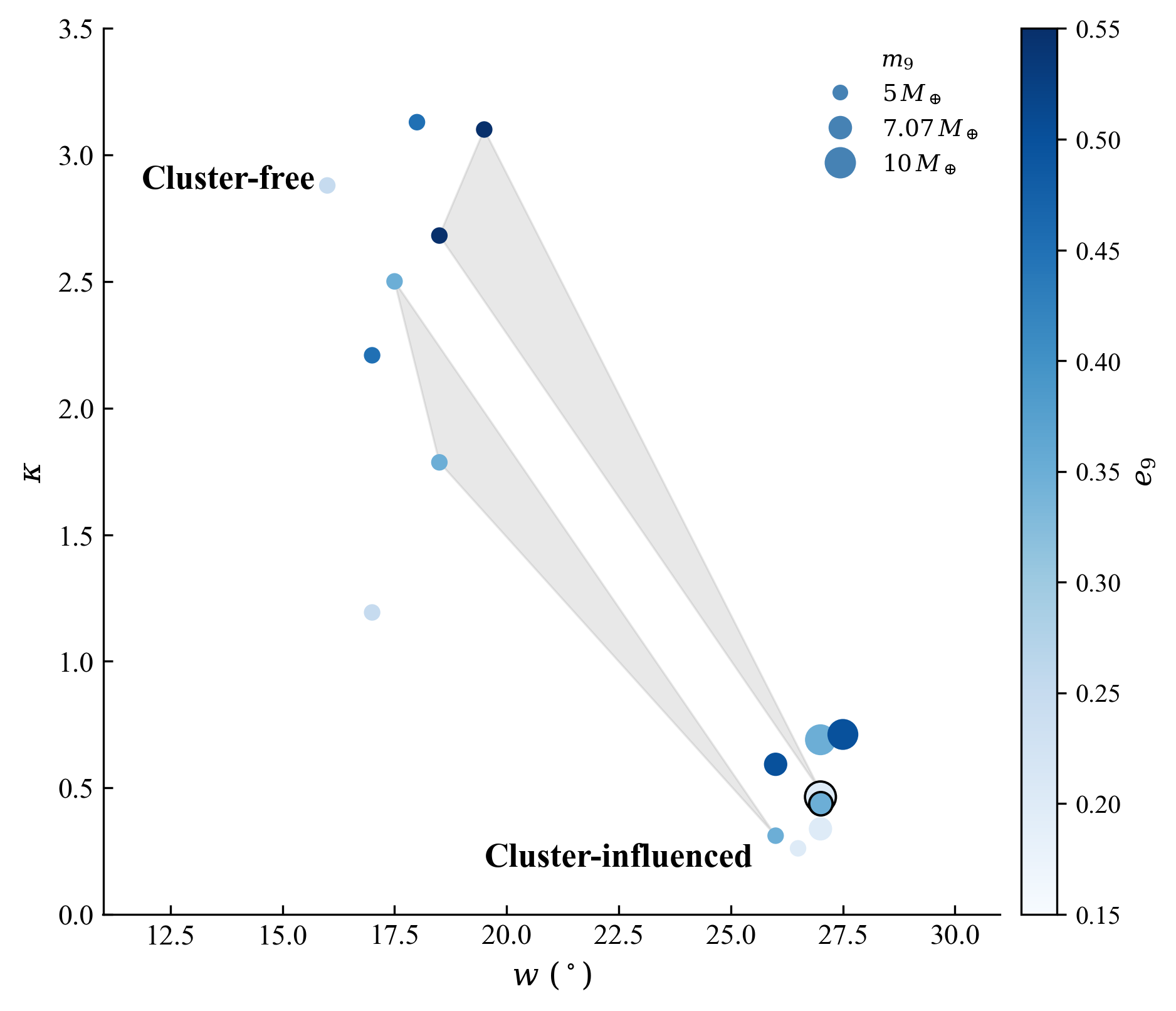}
\caption{Von-mises $\kappa$ versus inclination distribution width $w$ for Planet Nine simulations. Circle size indicates planet mass $m_9$ and color indicates eccentricity $e_9$. Gray line and shaded band connect cluster-influenced and cluster-free simulations sharing the same $m_9$ and similar $e_9$, illustrating the shift in both $w$ and $\kappa$ as a function of Planet Nine parameters.}
\label{fig:p9_clustering}
\end{figure}
The distant TNOs are known to exhibit perihelion clustering, so we check if any of our simulations can reproduce this. We quantify perihelion clustering using the von Mises concentration parameter $\kappa$, following a procedure similar to that used in \citet{GabK}, fit to the distribution of longitudes of perihelion $\varpi$ for TNOs with $a \geq 250 \, \mathrm{AU}$, $q \in (40, 100) \, \mathrm{AU}$, and $i \leq 40^\circ$. The von Mises distribution is the natural analogue of the Gaussian on the circle, with $\kappa = 0$ corresponding to a uniform distribution and larger $\kappa$ indicating stronger clustering. We fit $\kappa$ via maximum likelihood, solving $I_1(\kappa)/I_0(\kappa) = \bar{R}$, where $\bar{R} = |\langle e^{i\varpi} \rangle|$ is the mean resultant length of the angle distribution and $I_0$, $I_1$ are modified Bessel functions. The $a$ cut ensures that we largely focus on the subset of TNOs affected by Planet Nine, and the $q$ and $i$ cuts constitute crude proxies for observability. We find that cluster-free initial condition Planet Nine simulations can reproduce strong perihelion clustering ($\kappa > 1$), while cluster-influenced simulations produce weaker clustering ($\kappa < 1$, see Figure~\ref{fig:p9_clustering}).
\section{Discussion}\label{sec:Discussion}

\subsection{Implications and Future Work}
Under the assumptions that observational biases are insignificant and that the implantation of sednoids is dominated by the strongest stellar encounter, \citet{hu2025earlystellarflybysunlikely} showed that close stellar encounters are difficult to reconcile with the observed inclination distribution and alignment of the distant population of TNOs, suggesting that primordial cluster excitations were minimal. Meanwhile \citet{Nesvorny2023} showed that simulations with minimal cluster perturbations cannot reproduce the radially extended semi-major axis distribution of the scattered disk objects, pointing instead to strong cluster excitations. We explored whether this tension could be reconciled if the distant TNOs were initially excited by strong cluster perturbations and subsequently confined to observed levels by Planet Nine. We find that Planet Nine cannot confine the distant objects sufficiently. Thus, either the solar system experienced strong cluster perturbations and the low inclinations of the current census can largely be explained by observational biases, or it did not experience strong cluster perturbations.

One natural avenue for future work lies in exploring other ways to produce the radially extended distribution of the SDOs besides strong cluster perturbations. Planet Nine perturbations are a potential mechanism, although we note that simulations from  \citet{Nesvorny_2017} demonstrated that P9s with $m_9 \in [10,30]M_{\oplus}$ and $i_9 \in [0^\circ,30^\circ]$ could not change the radial extension of the SDOs. Nevertheless, dedicated simulations exploring the lower-mass Planet Nine parameter space are needed to explore whether Planet Nine can produce the radial extension of the SDOs from realistic initial conditions. Alternatively, a rogue planet may be able to produce the radial extension by scattering SDOs onto wider orbits before being ejected from the solar system \citep{Gladman_2006, huang2022rogue}. Again, much work is required to determine whether this is a viable scenario, including dedicated simulations to understand how the rogue planet evolves into the scattered disk, how it affects Neptune, and how it affects the KBOs.

\subsection{Caveats and Considerations}
\begin{itemize}
    \item We deliberately use initial conditions that incorporate a strong variant of cluster effects in order to study the scenario with the maximum possible cluster excitation within the constraints imposed by the cold classical belt. However it is possible that there is an intermediate variant of cluster effects that excites the distant objects to a lesser extent. We did not explore that scenario here because the semi-major axis distribution of the SDOs points to strong cluster perturbations identical to those used here. Nevertheless the intermediate cluster effects scenario is worth investigating.
    \item We infer the intrinsic inclination distribution of the distant objects from a sample of only 19 objects satisfying our selection criteria ($q \in (40, 80)$ AU, $a \in (200, 2000)$ AU). This small sample size introduces uncertainty in the inferred distribution width $w_{obs}$, since the likelihood function is broad near its peak when there is limited data, and many values of $w$ yield comparable likelihoods. We quantify this uncertainty by following the Monte Carlo calibration procedure described in \citet{Brown_2001}, wherein we calculate $D\sqrt{N}$ for $10^4$ instances of 19 inclinations drawn from the inferred distribution with $w_{obs} = 12^\circ$. By drawing 19 objects each time, our calibration procedure inherently accounts for small-sample statistics. Additionally, our cluster-influenced simulations produce inclination distributions with $w \approx 26^{\circ}$--$27.5^{\circ}$ (Table~\ref{tab:sim_sigma}) and indeed are visibly flat rather than sine-half-gaussian, so even though our modest sample size results in some uncertainty in the absolute value of $w_{obs}$, the coldness of the cluster-free simulations and the overexcited nature of the cluster-influenced ones is apparent. The Vera C. Rubin Observatory's Legacy Survey of Space and Time (LSST) is expected to increase the known TNO population by roughly an order of magnitude \citep{Ivezic_verarubin}, and the expanded sample of high-$q$ TNOs will tighten constraints on the intrinsic inclination width $w_{obs}$.
    \item The \citet{Brown_2001} debiasing method is derived under the assumption of circular orbits, which is clearly violated for high-$e$ TNOs. However, \citet{Brown_2001} tested the method's robustness by generating synthetic populations with known inclination distributions and eccentricities over $e \in [0,1]$, and then attempting to recover the input distribution using this method. Even for high eccentricity cases, the recovered inclination width was accurate to within 25\%, and fits were never rejected at the $2\sigma$ level. A 25\% systematic error on $w_{obs} \approx 12^{\circ}$ corresponds to $\sim 3^{\circ}$, far smaller than the $\sim 15^{\circ}$ gap between our measurement and the cluster-influenced predictions.
    \item We note that Table \ref{tab:sim_sigma} suggests some dependency between Planet Nine mass and $w$, further suggesting that there may be Planet Nine parameters that result in much lower $w$ consistent with the observed distant objects. We did not pursue this by exploring even higher Planet Nine masses because it has been shown that higher mass Planet Nine scenarios cannot simultaneously reproduce the observed apsidal and orbital-plane clustering of the distant TNOs (e.g., \citealt{batygin2019planet,brobat2021}). Our simulations span the range of masses and eccentricities typically invoked in Planet Nine studies, and the variation in $w$ across this parameter space is only $\sim 1.5^\circ$, far smaller than the $\sim 15^\circ$ gap between the cluster-influenced predictions and the observations. 
\end{itemize}
\section{Conclusion}\label{sec:Conclusion}
In this paper, we ran a suite of physically motivated N-body simulations of distant trans-Neptunian objects, initialized from a synthetic population obtained via numerical experiments of the early Solar System. We simulate the interactions of this population with the known giants, various realisations of Planet Nine, and the effects of the Galactic tide and passing stars over 4 Gyr. Our simulations demonstrate that Planet Nine is unable to confine the inclinations of the distant population to currently observed levels after they have been excited by strong cluster perturbations during the Solar System's early history. This rules out one proposed explanation for the concurrent observations of radially extended scattered disk objects and dynamically cold distant populations. Discoveries of addition TNOs by the Vera Rubin Observatory will enable us to further disentangle the relative roles of the birth cluster and subsequent encounters (and perhaps proposed unseen objects) in sculpting the outer solar system, as well to better reconstruct the dynamical history of the solar system as a whole.

\appendix
\section{Stellar Flyby Constraints from the Cold Classical Belt}

As outlined previously, another constraint on cluster perturbations of the early solar system (besides the high-q TNOs) comes from the cold classical belt. We briefly use numerical simulations to explore this constraint and place upper bounds on $\chi$ and the closest probably stellar flyby of the solar system.

The closest stellar flyby consistent with the observed TNOs is unknown. Estimates of the closest flyby pericenter consistent with the observed TNOs vary widely. It has been suggested that a moderate flyby with $q \approx 800 \, \mathrm{AU}$ could explain some features of the TNOs, such as the detached orbit of Sedna \citep{Morbidelli_2004}. It has also been argued that flybys with $q < 1000 \, \mathrm{AU}$ are statistically improbable within the context of cluster dynamics and solar migration models \citep{hu2025earlystellarflybysunlikely}. At the other extreme, numerical simulations have been used to advocate for encounters as close as $q = 110 \, \mathrm{AU}$ \citep{Pfalzner_2024}. Different arguments about the strength of stellar perturbations reflect varying assumptions about the Sun's cluster residence time and stellar density. Thus the debate about how strongly flybys sculpted the solar system is really a debate about what kind of environment the Sun was born in.

\citet{Batygin_2020} argued that the dynamically pristine cold classical Kuiper belt objects can be used as a constraint on any external perturbations of the solar system. The cold classical objects are almost certainly the outcome of in-situ formation \citep{batygin2011retention,nesvorny2019trans}. As argued by \citet{Morby_2004}, this population appears to have avoided the strong perturbations that afflicted other TNOs throughout their history. Several lines of evidence support this interpretation. Cold classicals exhibit low inclinations ($i < 5^\circ$) and eccentricities ($e < 0.1$) \citep{Gladman_2008}, inconsistent with the dynamical excitation produced by close encounters with Neptune. Pertinently, the free inclinations of the cold classicals follow a Rayleigh distribution with a scale parameter of $1.7^\circ$ \citep{Brown_2001}. They also host an unusually high fraction of wide binaries \citep{Stephens_2006}, which would have been readily disrupted by Neptune scattering \citep{Parker_2010}. Additionally, their uniformly red optical colors \citep{Trujillo_2002, Lykawka_2005} suggest a common and undisturbed formation environment, in contrast with the broader Kuiper belt's diverse composition. Because of their dynamical fragility and orbital isolation, any successful perturbation scenario must preserve their distinct orbital and physical properties. Arguing thus, \citet{Batygin_2020} modeled the cumulative effect of hyperbolic stellar flybys on this distribution using orbit-averaged secular perturbation theory, and found that encounters with $q < 240 \, \mathrm{AU}$ would excite the inclinations of the cold classicals beyond observed levels.

Here, we use a suite of $N$-body simulations to constrain the full range of stellar flyby geometries consistent with the survival of the cold classical Kuiper belt, while making minimal assumptions about the underlying encounter conditions. We also use our constraints to corroborate the $\chi$ upper bound reported in \citet{Batygin_2020}. Our work constitutes a more thorough version of the numerical experiment run in \citet{Siraj_2025}, with an enhanced assortment of integrations.

\section{Simulation Set-up}
Each simulation was initialized by placing the Sun and giant planets at their present-day orbital positions, using real-world ephemerides from the JPL Horizons database. All four giant planets are included explicitly as fully interacting massive bodies. This approach assumes that the current giant planet architecture was already in place at the time of the stellar flyby and does not attempt to model earlier dynamical processes such as planetary migration or the instability phases proposed in the Nice model. While this simplification omits aspects of early solar system evolution, it enables a controlled study of stellar encounters on the cold classical Kuiper belt under conservative and well-constrained conditions. Cold classical KBOs were likewise initialized at their current orbits using ephemerides from the Minor Planet Center. Objects were selected by filtering for perihelia in the range $42 \, \mathrm{AU} < q < 48 \, \mathrm{AU}$ and inclinations $i < 5^\circ$.

Each flyby star was initialized $20000 \, \mathrm{AU}$ from its perihelion. Its velocity at infinity was drawn from a Maxwell--Boltzmann distribution with a scale parameter of $\sqrt{2} \times 1 \, \mathrm{km \, s}^{-1}$. The stellar mass was drawn from the \citet{Kroupa_2001} initial mass function (IMF). Pericenters were drawn uniformly between $100$ and $1000 \, \mathrm{AU}$. The flyby's inclination, argument of periapsis, and longitude of ascending node were each sampled uniformly across $0^\circ$--$180^\circ$, $0^\circ$--$360^\circ$, and $0^\circ$--$360^\circ$ respectively. In total, $15000$ unique initial conditions were simulated.

Each system was numerically integrated from when the flyby star was $20000 \, \mathrm{AU}$ away from perihelion to when it was $3000 \, \mathrm{AU}$ past perihelion, using the \texttt{Mercurius} integrator and the \texttt{REBOUND} software package \citep{Rein_2012}.

\section{Upper Bound on $\chi$}

We decompose the inclinations of the cold-classical belt objects from the final snapshot of each simulation into free and forced components, and use a KS test to compare the free inclination distribution to a Rayleigh distribution with scale parameter $1.7^\circ$. If the p-value is less than $0.05$ or the scale parameter of the best fit Rayleigh distribution falls outside $1.7\pm0.1^\circ$, then the simulated cold belt is inconsistent with the observed cold belt, and we can say that the stellar flyby has 'destroyed' the pristine cold classical belt.

\begin{figure*}[t]
\centering
\includegraphics[width=0.48\textwidth]{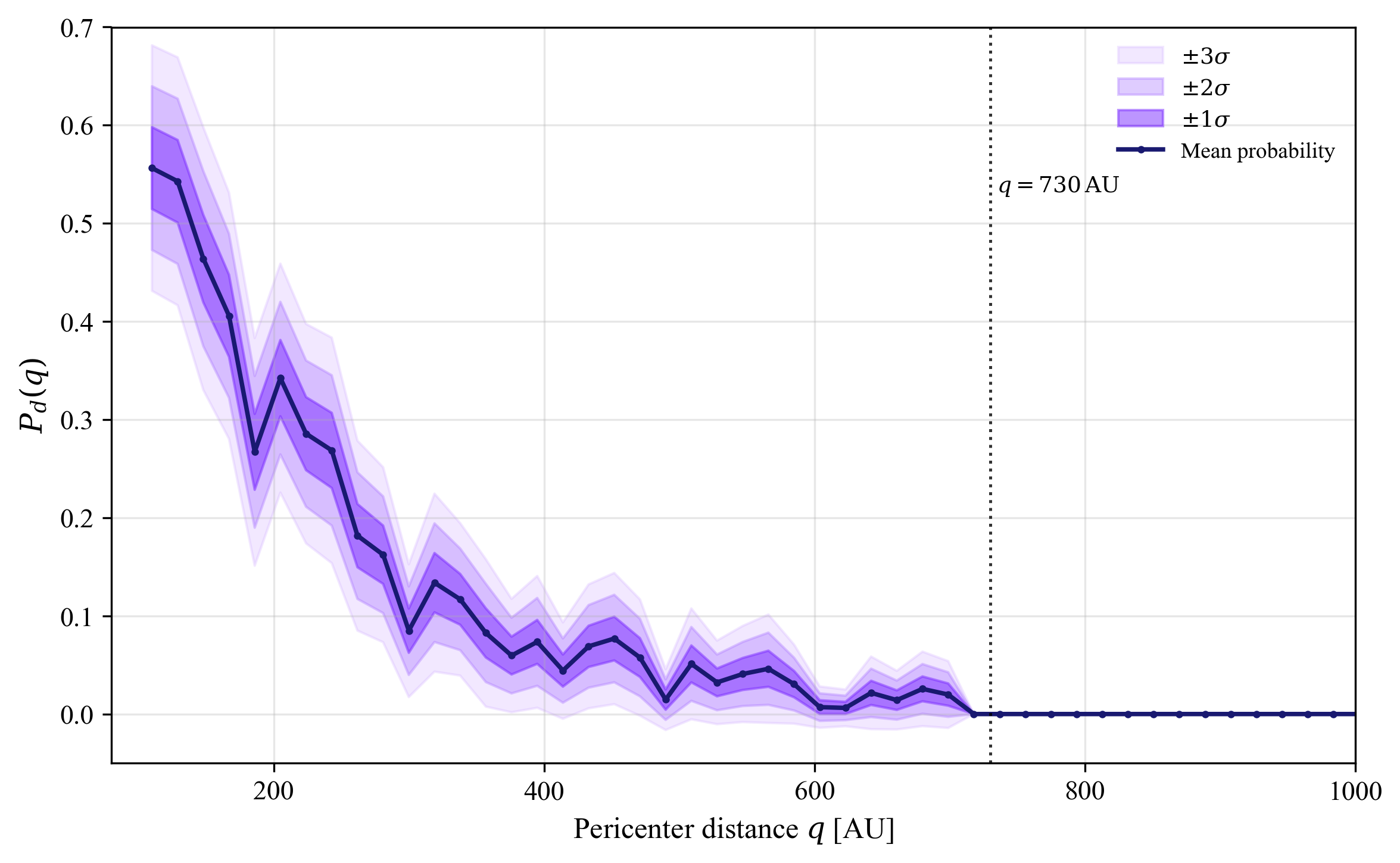}
\hfill
\includegraphics[width=0.48\textwidth]{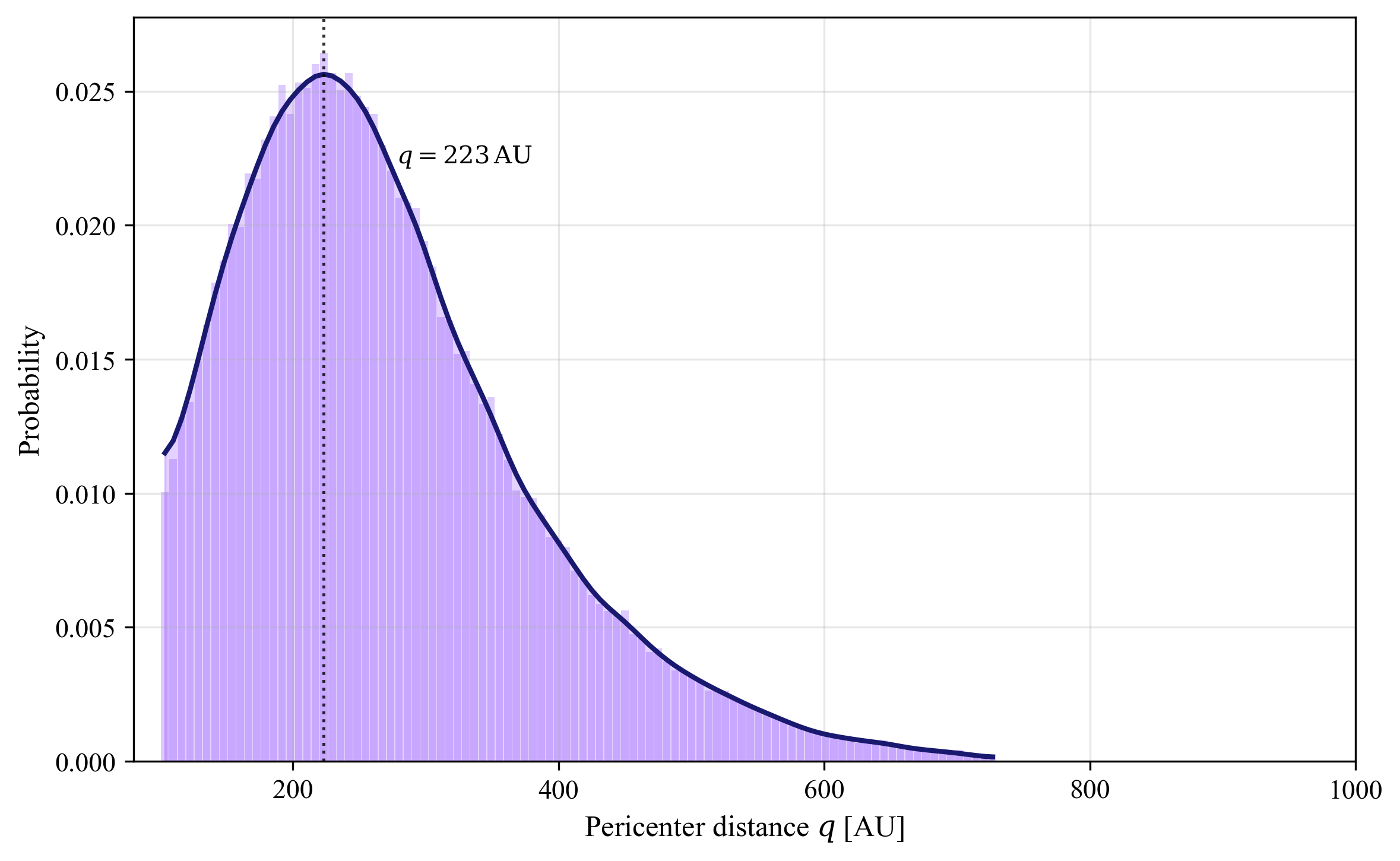}
\caption{Left: $P_d(q)$. Right: Probability density of the closest encounter pericenter in a destructive series of flybys, obtained from a Monte Carlo experiment. The distribution peaks at $q = 223\,\mathrm{AU}$.}
\label{fig:destruction_probability}
\end{figure*}

Let $N$ be the number density of stars and $v$ be the typical relative velocity of stars. Then the flux of stars (number of stars crossing a unit area per unit time) is $\Phi = Nv$. The number of encounters whose impact parameter is less than $b$ in a time interval $dt$ is given by
\begin{equation}
    dN(< b) = Nv \, 2\pi b \, db \, dt.
\end{equation}
The impact parameter $b$ is related to the pericenter $q$ by
\begin{equation}
    b = q \sqrt{1 + \frac{2GM_\odot}{q v_\infty^2}}.
\end{equation}
The effective cross-section for encounters with pericenter less than $q$ is therefore
\begin{equation}
    \sigma(<q) = \pi b^2 = \pi q^2 \left(1 + \frac{2GM_\odot}{q v_\infty^2}\right).
\end{equation}
Let $P_d(q)$ be the probability that the cold belt is destroyed by a flyby with pericenter $q$ (see Figure \ref{fig:destruction_probability}). Integrating over all pericenters, the total number of destructive flybys is:

\begin{equation}
    N_d(< q) = Nv\tau \int_{q_{\min}}^{q} P_d(q') \, 2\pi q' \left(1 + \frac{2GM_\odot}{q' v_\infty^2}\right) dq' \equiv Nv\tau \, A(q),
\end{equation}
where $A(q)$ is the cumulative effective cross-section for destructive encounters, including gravitational focusing. We note that while we treat $A$ as solely a function of $q$, there is in principle a secondary dependence on other flyby parameters such as stellar mass and orbital inclination. To quantify the relative importance of each parameter, we trained an XGBoost gradient-boosted classifier to distinguish Rayleigh distrubuted cold belts from non-Rayleigh ones as a function of variables describing the flyby star ($q$, $M_\star$, $v_\infty$, $i$, $\omega$, $\Omega$). Using gain-based feature importance scores, we find that $q$ is the dominant predictor of whether a flyby destroys the cold belt or not, with an importance score roughly twice that of the next most important feature, $M_\star$. The remaining parameters have negligible scores. This confirms that treating $A$ as a function of $q$ alone is reasonable. Qualitatively, the secondary mass dependence is intuitive, with more massive stars having higher destruction probabilities, for a given flyby geometry. A more detailed exploration of mass and other parameter dependencies is deferred to future work.

Since the cold belt is preserved, the expected number of destructive flybys should be less than $1$, so $N_d < 1$. We can use this to place an upper bound on $N\tau$:
\begin{equation}
    N\tau < \frac{1}{v \, A(q_{\mathrm{max}})} .
\end{equation}

From our simulations, the critical pericenter distance is $q_{\mathrm{max}} = 730 \, \mathrm{AU}$, beyond which $P_d \to 0$. 

Numerically integrating with the gravitational focusing correction for $v_\infty = \sqrt{2} \, \mathrm{km/s}$, the cumulative effective cross-section is
\begin{equation}
    A(q_{\mathrm{max}}) = 6.2 \times 10^{5} \, \mathrm{AU}^2 = 1.5 \times 10^{-5} \, \mathrm{pc}^2.
\end{equation}
Using $v = \sqrt{2} \, \mathrm{km/s} = 1.45 \, \mathrm{pc/Myr}$:
\begin{align*}
    N\tau &< \frac{1}{v \, A(q_{\mathrm{max}})}\\
    &= \frac{1}{(1.45 \, \mathrm{pc/Myr}})(1.5 \times 10^{-5} \, \mathrm{pc}^2) \\
    &= 4.7 \times 10^{4} \, \mathrm{Myr/pc}^3.
\end{align*}

This is consistent with the upper bound of $\sim 2-3 \times 10^4 \, \mathrm{Myr/pc}^3$ reported by \citet{Batygin_2020}.

While the $730 \, \mathrm{AU}$ pericenter distance represents the maximum for a single destructive flyby, a solar system born in a cluster likely experienced a series of stellar encounters rather than just one. To determine the most $\textit{probable}$, pericenter distance, we model these repeated encounters with a Monte Carlo experiment. The probability of a flyby with pericenter $q$ occurring scales with $\pi q^2$, so farther-out flybys occur more frequently. We pick $q$ from a distribution that goes with $q^2$, and uniformly draw a number from $[0,1]$. If the random number drawn is less than $P_d(q)$, then we say the belt is destroyed. Otherwise, we draw another $q$, repeat the process until the belt is destroyed, and record the smallest $q$ from that series of draws. We repeat this $100000$ times. The number of experiments in which some particular $q$ is the smallest pericenter in the destructive series, divided by the total number of experiments, gives us the probability that the closest encounter in a destructive series of encounters has pericenter $q$. We find that if the solar system was perturbed by a stellar flyby, the most probable pericenter for this encounter is $q = 223$ AU (see Figure \ref{fig:destruction_probability}). We note that our estimate is similar to a result derived by \citet{AdamsLaughline2001}. They ran $N$-body scattering experiments of passing stars encountering the solar system, computing the cross-section for these encounters to disrupt the orbits of the outer planets. Defining disruption as the excitation of the giant planets' eccentricities and inclinations beyond their observed values, they found a representative cross-section of $\langle\sigma\rangle \approx (400\,\mathrm{AU})^2$. We can equate this cross-section to $\pi b^2$, since a cross-section of area $\langle\sigma\rangle$ corresponds to a circular target of radius $b$ within which a passing star must approach for the encounter to be disruptive. This gives $b \approx 225\,\mathrm{AU}$, essentially identical to our most probable pericenter of $q = 223\,\mathrm{AU}$.


\end{document}